\documentclass[a4paper,11pt]{article}
\usepackage{pos}

\title{Tethered Balloons for Radio Detection of Neutrinos?}

\author*[a]{Rachel Scrandis}
\author[a]{Cosmin Deaconu}

\affiliation[a]{Kavli Institute for Cosmological Physics, Department of Physics, Department of Astronomy \& Astrophysics, University of Chicago,
Chicago, IL 60637}

\emailAdd{rscrandis@uchicago.edu}
\emailAdd{cozzyd@kicp.uchicago.edu}

\abstract{The long-duration balloon platform for radio detection of energetic neutrinos, pioneered by ANITA, affords large instantaneous effective areas but has limited livetime. Conversely, tethered balloons, traditionally used for radio detection of non-science objectives (such as electronic surveillance), allow for much longer livetimes, albeit at significantly lower altitudes. In this contribution, a tethered balloon platform for neutrino detection is considered, including estimates of the neutrino sensitivity and a discussion the feasibility of such a platform. Both the Askaryan and tau-neutrino induced Extensive Air Shower channels are explored as target science data for the platform. Tethered balloons locations on ice sheets, land, and in steep valleys or fjords will be considered. Ground-based array calibration and cosmic-ray air shower use cases will also be briefly commented on.}

\FullConference{10th International Workshop on Acoustic and Radio EeV Neutrino Detection Activities (ARENA2024)\\
11-14 June 2024\\
The Kavli Institute for Cosmological Physics, Chicago, IL, USA\\}

\begin{document}
\maketitle

\section{Introduction}
Astroparticle detectors searching for rare, ultrahigh-energy particles are designed to maximize exposure.  Experiments such as TA, Pierre Auger, Radio Neutrino Observatory in Greenland (RNO-G), IceCube, BEACON, and GRAND operate detector arrays over years to reach high enough statistics to detect the highest energy particles\citep{matthews2023results,aab2020measurement,aguilar2021design,icecube2023observation,Southall_2023,grandcollaboration2023giantradioarrayneutrino}. Alternatively, balloon experiments such as  ANITA, PUEO and EUSO-SPB rely on surveying a larger volume over a relatively short period of time to reach necessary sensitivities \citep{anita,eusospb,abarr_payload_2021}. An advantageous platform would leverage the strengths of both these approaches: the long exposure time of a ground array with the increased aperture of the elevated balloon platform. 

Tethered balloons, otherwise known as tethered aerostats, are blimp-shaped balloons connected to a mooring system (see Figure \ref{Aerostat_blueprint}), allowing the balloon to flying multiple kms in the air whilst being anchored to one spot \citep{AtlasLTA, Altaeros}. Historically these systems have been sparingly used in scientific cases. 

The Polar Stratospheric Telescope (POST) was a proposed near-IR and optical telescope planned to fly on a tethered aerostat in the Arctic and/or Antarctica \citep{bely1996post}. The science case eventually was fulfilled by the Stratospheric Observatory For Infrared Astronomy (SOFIA) experiment, which used a Boeing 747 as the platform \citep{gehrz2009new}. Challenges for POST included the stringent stability constraints on the telescope, which needed a stable FOV of at least 12 mas, a requirement that is much more relaxed for astroparticle applications. 

Using tethered balloons as an astroparticle detector platform is not a new idea either. The Radio Ice Tethered Antennae (RITA) experiment proposed attaching directional antenna to a tethered balloon and looking down at Antarctic ice at Vostok Station to see Askaryan signals, a design similar to ANITA but with longer exposure and lower flight altitude \cite{BESSON2012S50}. 

Currently, the most common use of tethered balloons in science is atmospheric monitoring. The Atmospheric Radiation Measurement (ARM) government facilities use small tethered balloons to measure atmospheric aerosol content, wind characteristics, and humidity levels at various altitudes \citep{ARM}. In this proceedings, we will discuss the modern capabilities of large tethered aerostats, potential astroparticle detector concepts, and challenges that exist with this platform.

\section{General Aspects of Tethered Aerostats}
\begin{figure}[tb]
\centering
        \includegraphics[width=.55\textwidth]{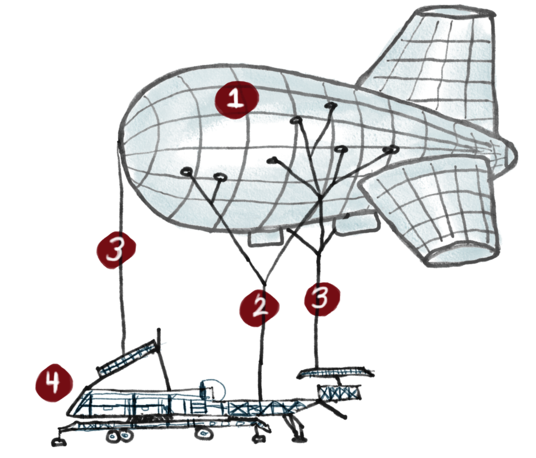}
        \caption{Typical Tethered Aerostat Platform \citep{AtlasLTA}. (1) Helium Filled Aerostat (2) Power Tether (3) Support Tethers (4) Ground Rig.}
        \label{Aerostat_blueprint}
\end{figure}
Tethered Aerostats have four main characteristics, as outlined in Figure \ref{Aerostat_blueprint}. The helium-filled balloon features fins that provide the aerostat lift and stability in the presence of wind. The underside of the balloon carries an electronics box, which houses connections to the power tether. The power tether also carries data and communications to the ground, which relaxes the onboard instrument requirements typical of normal weather balloon experiments. Additional anchor support is provided by physical tethers with no power or data lines. The ground mooring system is the main anchor point, capable of rotating around a central point to counteract wind forces. Modern mooring systems are capable of launching and bringing down aerostats autonomously, an important capability as the aerostats must come down in high winds and for helium refilling \citep{Altaeros}. 

Commercial aerostats typically fly between 300-600m above ground \citep{AtlasLTA, Altaeros}, however in research and governmental applications, balloons have flown 1.5-20km, with the price of the balloon rapidly increasing for those that fly beyond 5km \citep{AtlasLTA,ARM}.  The <5km aerostats can hold up to 300kg of weight, and can fly for 15-45 days at a time before refilling. The moderate height and relaunch capabilities means that tethered balloons could provide an attractive platform for astroparticle detectors; the long integration time and large detector volume combines the strengths of ground arrays and balloon experiments.
\subsection{Modern Use Cases}
Most applications of tethered aerostats comes from the commercial and government sectors. World Mobile is an internet provider focused on servicing previously unconnected parts of the world (including Zanzibar and Pakistan) \citep{WorldMobile}. Instead of erecting towers across countries, the company deploys tethered aerostats to supply connection to the surrounding area. This allows them to reach customers at a fraction of the cost to erect a tower themselves. The Federal Emergency Management Agency (FEMA) uses tethered balloons for similar purposes; as part of disaster relief, if communications infrastructure is destroyed, aerostats with communication capabilities can be deployed whilst first responders are servicing an area \citep{FEMA}. Given their semi-permanent set-up, such tethered aerostats can be used for long term disaster relief, and then be deconstructed and stored until the next use. The U.S. Customs and Border Protection’s Air and Marine Operations use aerostats for border surveillance. The Tethered Aerostat Radar System (TARS) is a fleet of radar equipped tethered balloons that fly along the coastal boarders of America \citep{TARS}. TARS is built to detect any low flying and/or smaller aircrafts, commonly used for illegal drug and human trafficking. Although making up 8\% of the radar fleet, TARS is responsible for catching 68\% of all stopped illegal transits.

All these modern use cases make use of the fact that all the balloon power can come from the ground, and the only onboard electronics need to be transmitting antennas and supporting hardware. The height provided by the balloons allow each application to service or monitor a large area on the ground with relatively few balloons. Lastly, each of these applications need near constant up time, with a tolerance for short interruptions as the balloon needs to come down. Thus, using aerostats as a low-power astroparticle detector is not a far extension from what they are currently being used for. The main difference between these applications and requirements on astroparticle detectors is a required low noise background. Exploration into mitigating noise emitted by the central mooring system and electronics within the payload will be needed before any detector can be effectively deployed. In the next sections, we will describe potential detector concepts that would benefit from the aerostat platform, all using the assumption that acceptable noise reduction steps have been taken.
\section{Askaryan Detector Concept}
A natural use for the tethered aerostat platform is as an UHE neutrino detector searching for the Askaryan emissions arising in ice. Much like ANITA or PUEO, an aerostat flying above Antarctica or Greenland would present a large instantaneous effective volume, making it a competitive detector in a few months integration time \citep{abarr_payload_2021}. A tethered balloon would fly lower than its long-duration balloon counterparts, likely at $\sim$few km, but could fly continuously for weeks. A lower altitude is also not necessarily a negative; by being closer to the ground, the balloon could detect lower-energy neutrinos than possible for ANITA and PUEO. For the detector itself, instead of hanging antennas below the payload (as done on long-duration balloons), conductive fabric could be sewn to the sides of the aerostat itself, making the balloon its own detector (see PUEO's Low Frequency Instrument antenna design for an example of such an antenna). The on-board triggering capabilities of a tethered balloon detector could also be relaxed, as the data line to the ground would provide access to a more robust DAQ system. Lastly, a tethered aerostat detector would also remain (nearly) stationary, meaning the background noise rejection (especially anthropogenic radio emission, which is the dominant background) is a great deal simpler compared to a moving balloon payload which travels large distances through various noise environments.

\begin{figure}[htp]
    \centering
    \includegraphics[width=.9\textwidth]{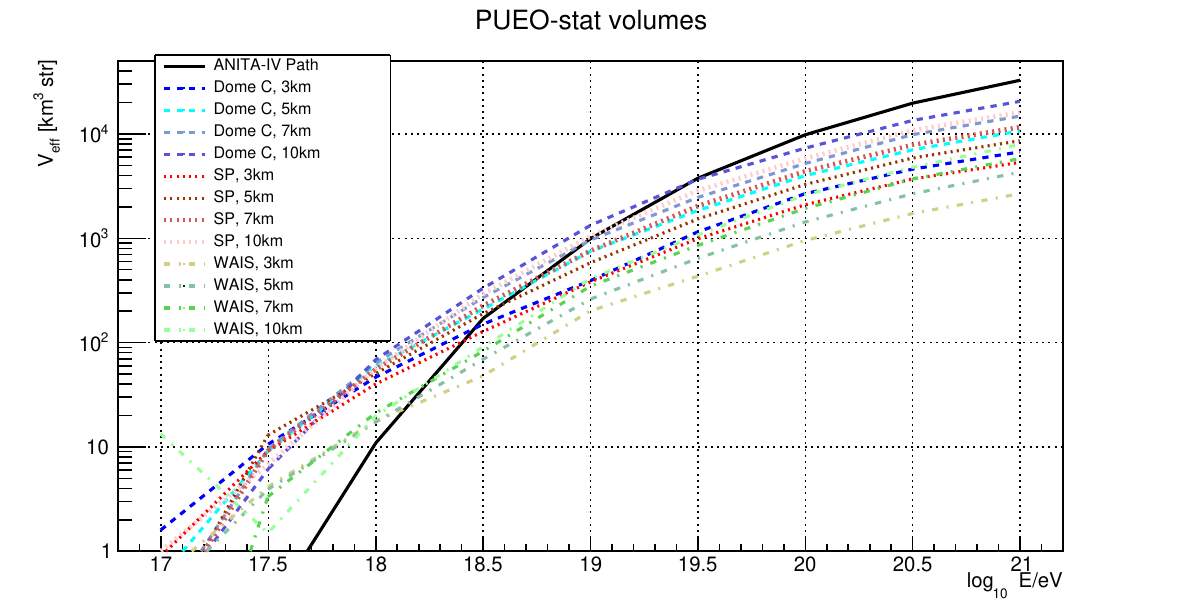}
    \caption{Rough projected effective volumes for a tethered aerostat UHE neutrino detector at various locations in Antarctica. Curves are compared to the same trigger on a ANITA-IV flight path; general characteristics of the lines should be considered rather than their absolute values.}
    \label{pueostat}
\end{figure}

Considering these strengths, the PUEO collaboration previously investigated using a tethered aerostat as their balloon platform. Using a simple Monte-Carlo simulation, they generated rough effective volumes estimates at various neutrino energies and balloon sites and compared them to previous ANITA flights; the results are seen in Figure \ref{pueostat}. These simulations assumed a trigger threshold 2.5x lower than ANITA-IV, which has been realized (and more) by the current PUEO trigger. Further refining could be done to improve the trigger further when considering the constant noise background that a tethered aerostat would encounter.
Notably, tethered aerostats provide increased performance at lower energies due to their closer proximity to the ground. At energies between $10^{18}$-$10^{20}$ eV, aerostats continue to be competitive with ANITA-IV. Thus, when considering that aerostats provide much longer livetimes, the overall sensitivity of such a balloon could quickly outpace the long-duration balloon platform. The biggest drawback (in addition to the comparative lack of the successful heritage for scientific purposes) is that tethered aerostats that can fly between 3-10km are much more expensive than a single-use long-duration balloon and likely require additional staffing.  A detailed design study would be necessary to properly compare costs and benefits.

\section{Tau Decay Detector Concept}
Tethered aerostats could be used to search for UHE neutrinos without leveraging large volumes of ice as well. Detectors such as BEACON and GRAND look specifically for tau air shower decays arising from $\nu_{\tau}$ CC interactions near Earth's surface. An aerostat, instead of relying on a mountainside to provide the necessary elevation to look down (as is done with BEACON and GRAND), could float above a large amount of land and search for this signal channel. Additionally, instead of being deployed on a remote mountaintop that could be RFI contaminated, an aerostat could be deployed at the base of a valley or fjord and use the walls of the depression as detector volume as well. This has the added benefit that the natural walls will act as insulation from RFI beyond the ridgeline. This detector concept can use conductive fabric antennas sewn onto the balloon fabric and ground based triggering schemes as well. Lastly, deployment in a valley or fjord represents a deployment site a great deal more accessible than that of an polar ice sheet counterpart.

Considering the quieter detector location and larger detector volume from the earth walls, a tethered aerostat tau decay experiment presents a great deal of potential. In a brief exploration using GRAND's simulation tools, placing a detector in a Nevada valley presents a factor of two increase in detector sensitivity just from the geographical location alone. Further investigations must be done to understand and optimize valley width and height against neutrino sensitivity.
\section{Small Aerostat Applications}
On the smaller scale (carrying capacity $<$20kg), tethered aerostats offer alternative calibration solutions for extended ground arrays. For such experiments, air shower calibration typically occurs using a drone carrying a pulser. Such drones can be expensive, RFI-emitting, only able to fly for a short time, and require pilot's licenses to operate. Small tethered balloons can fly up to a couple km high for hours while holding a pulser, and by outfitting the balloon with a GPS antenna, its exact location can be known. These aerostats also cost between \$1,000 - \$4,000, making them a cost effective alternative. Stationing these balloons across an array could also provide in-situ atmospheric monitoring. For dispersed arrays like BEACON or GRAND, such measurements could reduce systematic error between stations at different locations and environmental conditions. 

\bibliographystyle{JHEP}
\bibliography{Refs}

\providecommand{\href}[2]{#2}\begingroup\raggedright\begin{thebibliography}{10}

\bibitem{matthews2023results}
J.~Matthews et~al., \emph{Results from the telescope array},  in \emph{Journal of Physics: Conference Series}, vol.~2429, p.~012011, IOP Publishing, 2023.

\bibitem{aab2020measurement}
A.~Aab et~al., \emph{Measurement of the cosmic-ray energy spectrum above 2.5$\times$ 10 18 ev using the pierre auger observatory}, {\emph{Physical Review D} {\bfseries 102} (2020) 062005}.

\bibitem{aguilar2021design}
J.~Aguilar et~al., \emph{Design and sensitivity of the radio neutrino observatory in greenland (rno-g)}, {\emph{Journal of Instrumentation} {\bfseries 16} (2021) P03025}.

\bibitem{icecube2023observation}
R.~Abbasi et~al., \emph{Observation of high-energy neutrinos from the galactic plane}, {\emph{Science} {\bfseries 380} (2023) 1338}.

\bibitem{Southall_2023}
D.~Southall et~al., \emph{Design and initial performance of the prototype for the beacon instrument for detection of ultrahigh energy particles}, \href{https://doi.org/10.1016/j.nima.2022.167889}{\emph{Nuclear Instruments and Methods in Physics Research Section A: Accelerators, Spectrometers, Detectors and Associated Equipment} {\bfseries 1048} (2023) 167889}.

\bibitem{grandcollaboration2023giantradioarrayneutrino}
{GRAND Collaboration}, \emph{The giant radio array for neutrino detection (grand) collaboration -- contributions to the 38th international cosmic ray conference (icrc 2023)},  \href{https://arxiv.org/abs/2308.00120}{https://arxiv.org/abs/2308.00120}, 2023.

\bibitem{anita}
{\scshape ANITA} collaboration, \emph{{Constraints on the ultrahigh-energy cosmic neutrino flux from the fourth flight of ANITA}}, \href{https://doi.org/10.1103/PhysRevD.99.122001}{\emph{Phys. Rev. D} {\bfseries 99} (2019) 122001} [\href{https://arxiv.org/abs/1902.04005}{{\ttfamily 1902.04005}}].

\bibitem{eusospb}
{\scshape JEM-EUSO} collaboration, \emph{{EUSO-SPB1 mission and science}}, \href{https://doi.org/10.1016/j.astropartphys.2023.102891}{\emph{Astropart. Phys.} {\bfseries 154} (2024) 102891} [\href{https://arxiv.org/abs/2401.06525}{{\ttfamily 2401.06525}}].

\bibitem{abarr_payload_2021}
Q.~Abarr et~al., \emph{The {Payload} for {Ultrahigh} {Energy} {Observations} ({PUEO}): {A} {White} {Paper}}, \href{https://doi.org/10.1088/1748-0221/16/08/P08035}{\emph{Journal of Instrumentation} {\bfseries 16} (2021) P08035}.

\bibitem{AtlasLTA}
{Atlas LTA}, \emph{Our solutions: Communications and surveillance},  \href{https://atlas-lta.com/communications-surveilance/}{https://atlas-lta.com/communications-surveilance/}, 2020.

\bibitem{Altaeros}
Altaeros, \emph{Commercial solutions},  \href{https://www.altaeros.com/commercial/}{https://www.altaeros.com/commercial/}, 2024.

\bibitem{bely1996post}
P.Y.~Bely, H.C.~Ford, R.~Burg, L.~Petro, R.~White and J.~Bally, \emph{Post: Polar stratospheric telescope}, {\emph{Infrared and Submillimeter Space Missions in the Coming Decade: Programmes, Programmatics, and Technology} (1996) 101}.

\bibitem{gehrz2009new}
R.D.~Gehrz, E.E.~Becklin, I.~de~Pater, D.F.~Lester, T.L.~Roellig and C.E.~Woodward, \emph{A new window on the cosmos: The stratospheric observatory for infrared astronomy (sofia)}, {\emph{Advances in Space Research} {\bfseries 44} (2009) 413}.

\bibitem{BESSON2012S50}
D.~Besson, R.~Dagkesamanskii, E.~Kravchenko, I.~Kravchenko and I.~Zheleznykh, \emph{Tethered balloons for radio detection of ultra high energy cosmic neutrinos in antarctica}, \href{https://doi.org/https://doi.org/10.1016/j.nima.2010.11.046}{\emph{Nuclear Instruments and Methods in Physics Research Section A: Accelerators, Spectrometers, Detectors and Associated Equipment} {\bfseries 662} (2012) S50}.

\bibitem{ARM}
ARM, \emph{Tethered balloon systems},  \href{https://www.arm.gov/capabilities/instruments/tbs}{https://www.arm.gov/capabilities/instruments/tbs}, 2024.

\bibitem{WorldMobile}
{World Mobile}, \emph{How world mobile’s aerostats work},  \href{https://worldmobile.io/blog/post/how-aerostats-work}{https://worldmobile.io/blog/post/how-aerostats-work}, 2022.

\bibitem{FEMA}
FEMA, \emph{03oe-07-stat - aerostat, tethered (balloon)},  \href{https://www.fema.gov/grants/tools/authorized-equipment-list/03oe-07-stat}{https://www.fema.gov/grants/tools/authorized-equipment-list/03oe-07-stat}, 2024.

\bibitem{TARS}
{U.S. Customs and Border Protection}, \emph{Air and Marine Operations Tethered Aerostat Radar System (TARS)}, 2024.

\end{thebibliography}\endgroup

\end{document}